\documentclass[aps,prapplied,
amsmath,amssymb,reprint,longbibliography]{revtex4-1}

\usepackage{graphicx}
\usepackage{bm}
\usepackage{fancyref}
\usepackage{soul}

\usepackage[version=3]{mhchem}
\usepackage{xcolor}
\usepackage{titlesec}
\titlespacing*{\section}{0pt}{1.1\baselineskip}{\baselineskip}

\begin{document}

	\title[]{Quantum paraelectricity probed by superconducting resonators}
	
	\author{D. \surname{Davidovikj}\normalfont\textsuperscript{\ddag}}
	\email{d.davidovikj@tudelft.nl}
	
	\author{N. \surname{Manca}\normalfont\textsuperscript{\ddag}}
	\email{n.manca@tudelft.nl}
	\thanks{\\\ddag Authors equally contributed}
	\author{H. S. J. \surname{van der Zant}}
	\author{A. D. \surname{Caviglia}}
	\author{G. A. \surname{Steele}}
	\noaffiliation
	\affiliation{Kavli Institute of Nanoscience, Delft University of Technology, P.O. Box 5046, 2600 GA Delft, The Netherlands}
	\date{\today}
	
	\begin{abstract} %max 150 words
		Superconducting coplanar waveguide (CPW), resonators are powerful and versatile tools used in areas ranging from radiation detection to circuit quantum electrodynamics. Their potential for low intrinsic losses makes them attractive as sensitive probes of electronic properties of bulk materials and thin films. Here we use superconducting MoRe CPW resonators, to investigate the high-frequency (up to 0.3 GHz) and low temperature (down to 3.5 K) permittivity of \ce{SrTiO3}, a non-linear dielectric on the verge of a ferroelectric transition (quantum paraelectricity). We perform a quantitative analysis of its dielectric properties as a function of external dc bias (up to $\pm~\mathrm{15\,V}$), rf~power and mode number and discuss our results within the framework of the most recent theoretical models. We also discuss the origin of a fatigue effect that reduces the tunability of the dielectric constant of \ce{SrTiO3}, which we relate to the presence of oxygen vacancies.
	\end{abstract}
	
	\maketitle
	
	\section{INTRODUCTION} 
	
	\indent\indent In recent years, coplanar waveguide (CPW) resonators have proven to be a unique tool for probing a wide variety of excitations in circuits and materials, including superconducting qubits~\cite{Wallraff2004}, ferromagnetic spin ensembles~\cite{Schuster2010} and magnons~\cite{Huebl2013}. Due to their ultra-low ohmic losses, they usually exhibit high quality factors, making them very sensitive to external perturbations. As their properties are, to a large extent, defined by their geometry and by the dielectric response of their environment, they make a promising candidate as probes for materials with exotic electronic properties.
	Transition metal oxides (TMOs) heterointerfaces exhibit a variety of electronic and structural properties, including 2D superconductivity~\cite{Reyren2007}, ferroelectric/magnetic orders~\cite{Noheda2014} and negative capacitance~\cite{Zubko2016}.
	Probably the most ubiquitous representative of transition metal oxides is \ce{SrTiO3}. 
	Its chemical stability, lattice constant and dielectric characteristics make it one of the standard substrates for the growth of high-quality crystalline thin films.
	
	At low temperatures \ce{SrTiO3} is characterized by an extremely high dielectric response ($\varepsilon_r=23,000$ at 4K~\cite{Sakudo1971, Neville1972, Krupka1994,Geyer2005}) originating from an incipient ferroelectric transition (quantum paraelectricity). The permittivity is affected by electric fields, and below 4 K it can be lowered by more than one order of magnitude for a field of about 2 MV/m~\cite{Hemberger1995, Antons2005b}. The gate-tunability of the dielectric constant of \ce{SrTiO3}, particularly in thin-film form, has been used in applications of resonators incorporating \ce{SrTiO3} for voltage-controlled microwave filters~\cite{Galt1993, Findikoglu1995, Findikoglu1996, Sok1997, Fuke2000, Adam2002}. 
	 Beyond applications, one can also envision using superconducting resonators for sensing and exploring the dielectric properties of \ce{SrTiO3} itself, similar to the application of superconducting cavities to probe qubits, spin ensembles, and magnetic excitations. The integration of CPW technology with oxide heterostructures is a novel approach towards studying these complex materials, as a high-precision and device-oriented technique. However, it first requires understanding of how the presence of the \ce{SrTiO3} substrate affects the response of planar devices working at microwave frequencies.
	
	Here, we use superconducting CPW resonators fabricated on single-crystal \ce{SrTiO3} to explore the dielectric response of this quantum paraelectric material. As a reference, we fabricate a nominally identical resonator on single-crystal \ce{Al2O3} (sapphire). Since sapphire is a well-known standard substrate, with low dielectric losses~\cite{Blair1982}, we were able to perform a quantitative analysis on the dielectric response of the resonator fabricated on top of the \ce{SrTiO3} sample under different excitations by direct comparison of the two resonators.
	
	\section{EXPERIMENTAL SETUP}

\begin{figure}[ht]
			\includegraphics[width=1.0\linewidth]{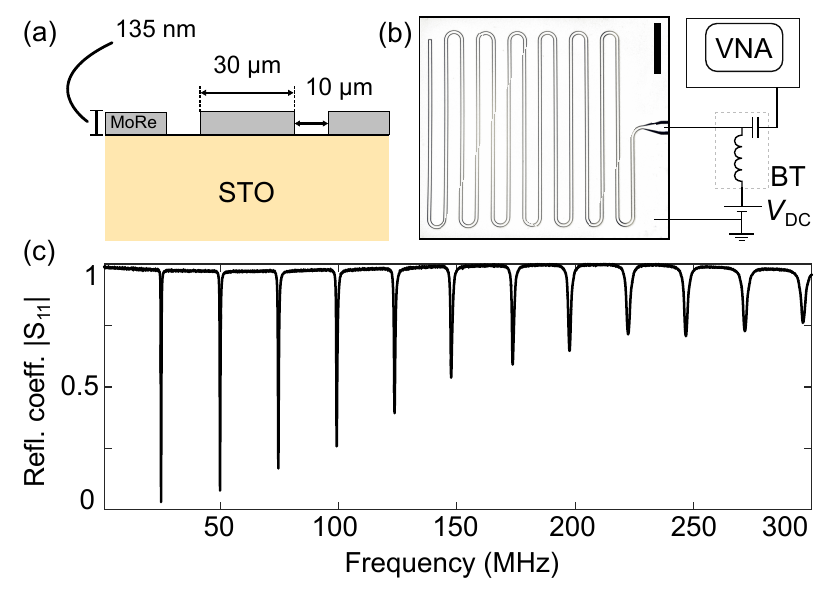}
			\caption{\label{fig:Fig1} Design of a superconducting CPW resonator as a probe of the dielectric response of \ce{SrTiO3}.
				(a) Cross-sectional schematic of the device.
				(b) Top-view optical image of the $\mathrm{\lambda /2}$ CPW resonator with a schematic of the measurement setup. (VNA: Vector Network Analyzer; BT: Bias tee)
				(c) Reflection measurement $\mathrm{\vert{S_{11}}\vert}$ taken at 3.5 K. 12 resonance modes are visible in the 1-300 MHz range.}
		\end{figure}

	\indent\indent A superconducting coplanar waveguide resonator is fabricated on top of a \ce{TiO2}-terminated single-crystal \ce{SrTiO3}(001) substrate ($5 \times 5 \times 0.5 \,\mathrm{mm^3} $). The CPW resonator is realized by a standard lithographic process followed by sputtering of a 135 nm thick molybdenum-rhenium (\ce{Mo_{0.60}Re_{0.40}}) film and lift-off in acetone. A cross-sectional schematic of the device is presented in Fig.~\ref{fig:Fig1}(a). The CPW resonator, is a 45~mm long line with a meander shape, the central strip is 30~$\mathrm{\mu m}$ wide, with a 10~$\mathrm{\mu m}$ gap to the ground plane.
	The geometric parameters are chosen such that a significant impedance mismatch is created between the transmission line ($\mathrm{Z_0 = 50 \,\Omega}$) and the resonator ($\mathrm{Z_{CPW} = 1 \,\Omega}$ at 3.5 K, which acts as a mirror for microwave photons). This, together with the open end of the resonator, creates a $\mathrm{\lambda/2}$ cavity. Fig.~\ref{fig:Fig1}(b) shows a top-view optical image of the device together with a schematic of the measurement setup. Our measurement scheme provides galvanic access to the central strip, so a dc bias can be easily applied with respect to the ground plane. Reflection measurements ($\mathrm{S_{11}}$) are performed using a vector network analyzer, while the sample is always kept at a fixed temperature of 3.5 K.
	
	We obtain information on the dielectric properties of the \ce{SrTiO3} by direct comparison with an identical resonator, fabricated on top of a sapphire substrate (see Supplemental Material S1). As the inductance of both waveguides is the same, the ratio of the frequencies of the fundamental modes of the two resonators ($f_1^\mathrm{Al_2O_3}$ and $f_1^\mathrm{STO}$) allows us to extract the effective dielectric constant of the \ce{SrTiO3} as follows:
	\begin{equation}
	\label{eq:varepsilon_vs_sapphire}
	\varepsilon_{r,eff}^\mathrm{STO} = \varepsilon_{r,eff}^\mathrm{Al_2O_3} \cdot \left({f_1^\mathrm{Al_2O_3}}/{f_1^\mathrm{STO}}\right)^2.
	\end{equation}
	
	The effective dielectric constant accounts for the electric field distribution through the vacuum above the substrate using a linear correction factor ($\varepsilon_{r,eff} = \alpha \cdot\varepsilon_r$). For dielectric constants in the range of the measured values for \ce{SrTiO3} ($\varepsilon_r$ = 2.000 - 30.000) the correction factor $\alpha^\mathrm{STO}$ is calculated to be $\alpha^{STO}_\mathrm{\varepsilon_r = 2,000} = 0.5000$ and $\alpha^{STO}_\mathrm{\varepsilon_r = 30,000} = 0.4998$ (using analytical expressions from~\cite{Simons2004}). Since sapphire has a much lower dielectric constant which is also anisotropic, $\alpha^\mathrm{Al_2O_3}$ is calculated using a finite elements simulation and is found to be $\alpha^\mathrm{Al_2O_3} = $ 0.6004 (see Supplemental Material S2).
	The dielectric constant of \ce{SrTiO3} is strongly anisotropic, where the different crystal orientations show a variation of $\varepsilon_\mathrm{r}$ higher than a factor of two~\cite{Sakudo1971}. This means that planar geometries will sense an effective dielectric constant arising from an average over the different crystal orientations, domain configurations and regions with different intensity of electric field (see Supplemental Material  S2).
		
	A reflection measurement $\mathrm{\vert{S_{11}}\vert}$ of the cavity modes is presented in Fig.~1(c), where twelve resonance modes are observed in the range from 1 to 300 MHz. The fundamental resonance frequency is 25.04 MHz, corresponding to $\mathrm{\varepsilon_r =}$ 27,400, calculated from the observed frequency of the sapphire sample $f_1^\mathrm{Al_2O_3} =$ 1.241 GHz.
	
	\section{ELECTRIC-FIELD DEPENDENCE OF  $\varepsilon_r$}
	
	\indent\indent The galvanic connection of the feedline to the CPW resonator enables us to apply a voltage difference between the central strip and the ground plane. The narrow 10~$\mathrm{\mu m}$ gap converts a relatively small voltage to an intense and localized average electric field (1 V $\approx$ 100 kV/m), whose magnitude is further enhanced in the proximity of the CPW by the inhomogeneous density of electric filed lines (see Supplemental Material S2).
	A similar local gating scheme was used recently in transport experiments in oxide systems~\cite{Stornaiuolo2014, Liu2015c, Monteiro2017}. The modification of the local dielectric constant caused by the dc bias results in a change of the total capacitance of the resonator and a corresponding shift of its resonance frequencies. By tracking the resonance frequency as a function of applied voltage across the gap, we probe the voltage dependence of the dielectric properties of \ce{SrTiO3} in a wide range of values.
	
\begin{figure}[]
		\includegraphics[width=1.0\linewidth]{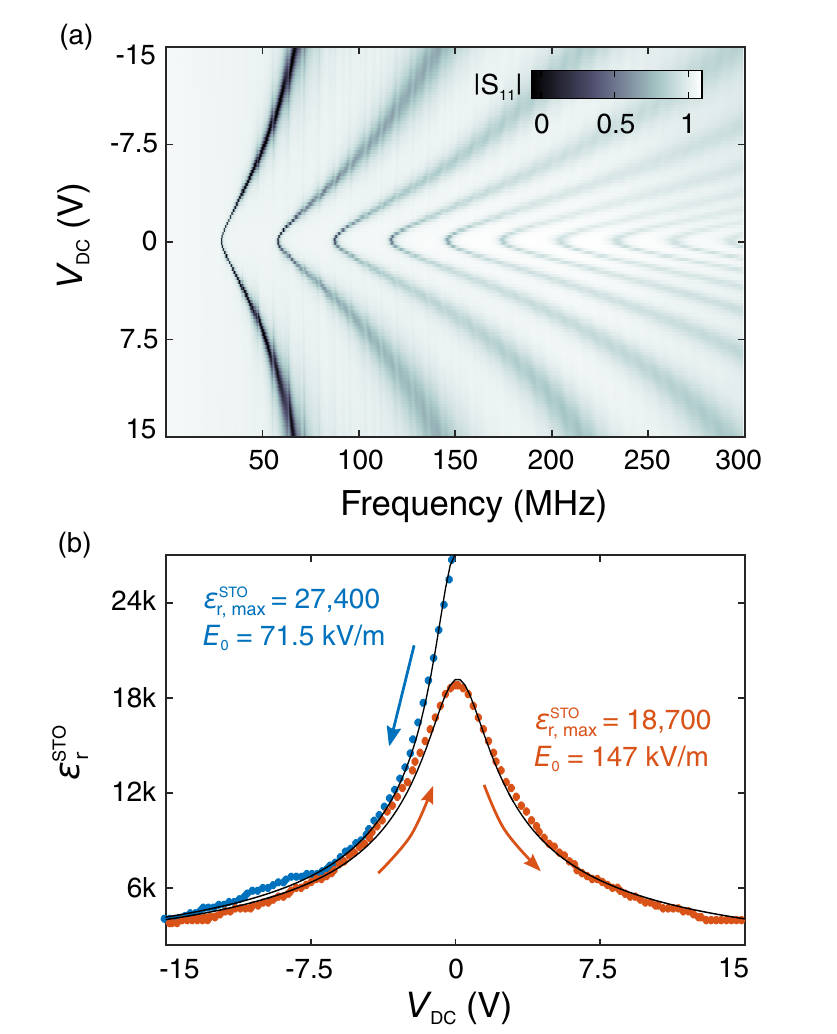}
		\caption{\label{fig:Fig2}
			dc voltage response of the resonator.
			(a) Color-plot of $\mathrm{|S_{11}|}$, showing the frequency shift of the cavity modes upon applying dc voltage in the $\pm$ 15 V range.
			(b) Voltage dependence of $\varepsilon_r$ of the \ce{SrTiO3} substrate calculated from the frequency shift of the first mode of the CPW resonator. The blue data points are measured right after cooldown while the red data points are taken after subsequent voltage sweeps. The black lines are fits to the data using eq. \ref{eq:eps_field}.}
	\end{figure}

	In Fig.~\ref{fig:Fig2}(a) we show a color plot of the reflection parameter $\mathrm{|S_{11}|}$ as a function of dc voltage from -15 to 15 V. By applying -15 V to the central strip, the fundamental mode shifts from 25 MHz to 66 MHz, which corresponds to a tunability of 160 \%. By applying 30 V the fundamental frequency shifts to 90 MHz, resulting in a tunability of 260 \% (see Supplemental Material S3).
	The extracted dielectric constant as a function of the applied voltage is plotted in Fig.~\ref{fig:Fig2}(b). We calculated $\varepsilon_r^\mathrm{STO}$ during the first voltage sweep just after the cooldown (black dashed line) and after subsequent sweeps in the $\pm$ 15 V range (orange solid line).  There is an evident non-reversible modification of the dielectric constant at low bias after the application of the voltage, where $\varepsilon_r^\mathrm{STO} (0)$ is reduced from the initial value of 27,400 to less than 19,000. This behavior, typically observed in \ce{SrTiO3}-based devices, is usually attributed to pinning of domain walls~\cite{Honig2013a}.
	The electric field dependence of the dielectric constant of \ce{SrTiO3} at low temperatures can be described in first approximation by the Landau-Ginzburg-Devonshire theory~\cite{Landau1984,Ang2004}:
	\begin{equation}
	\label{eq:eps_field}
	\varepsilon_r^\mathrm{STO} (E) = 1 + \frac{\varepsilon_r^\mathrm{STO} (0)}{[1+(E/E_0)^2]^{1/3}},
	\end{equation}
	where $E$ is the electric field and $E_0$ is a parameter related to the tunability of the dielectric constant with electric field. During the first field sweep (black dashed line), we extracted a value for $E_0$ = 71.5 kV/m, whereas in subsequent sweeps (orange line), this value increased to $E_0$ = 147 kV/m, indicating a decrease in the polarizability of the \ce{SrTiO3}. 	In Section V, we provide a detailed analysis of this effect. We note that both these values are about one order of magnitude lower than what was reported in literature from measurements on bulk single-crystal \ce{SrTiO3} with homogeneous electric fields ~\cite{Vendik1999}. This is due to the inhomogeneous distribution of the electric field, which is enhanced in the proximity of the central strip and the ground plane of the resonator. This allows us to affect the dielectric response at much lower voltages than the ones required in a double-capacitor geometry. It is worth mentioning, however, that despite of the fact that we are probing the volume average of the properties of a strongly non-linear material, the theoretical functional formulation still holds, supporting our analysis based on effective quantities.

	\begin{figure*}[ht]
	\begin{minipage}[c]{\textwidth}
		{
			\begin{minipage}[t]{0.75\textwidth}
				\mbox{}\\[-\baselineskip]
				\includegraphics[width=\linewidth]{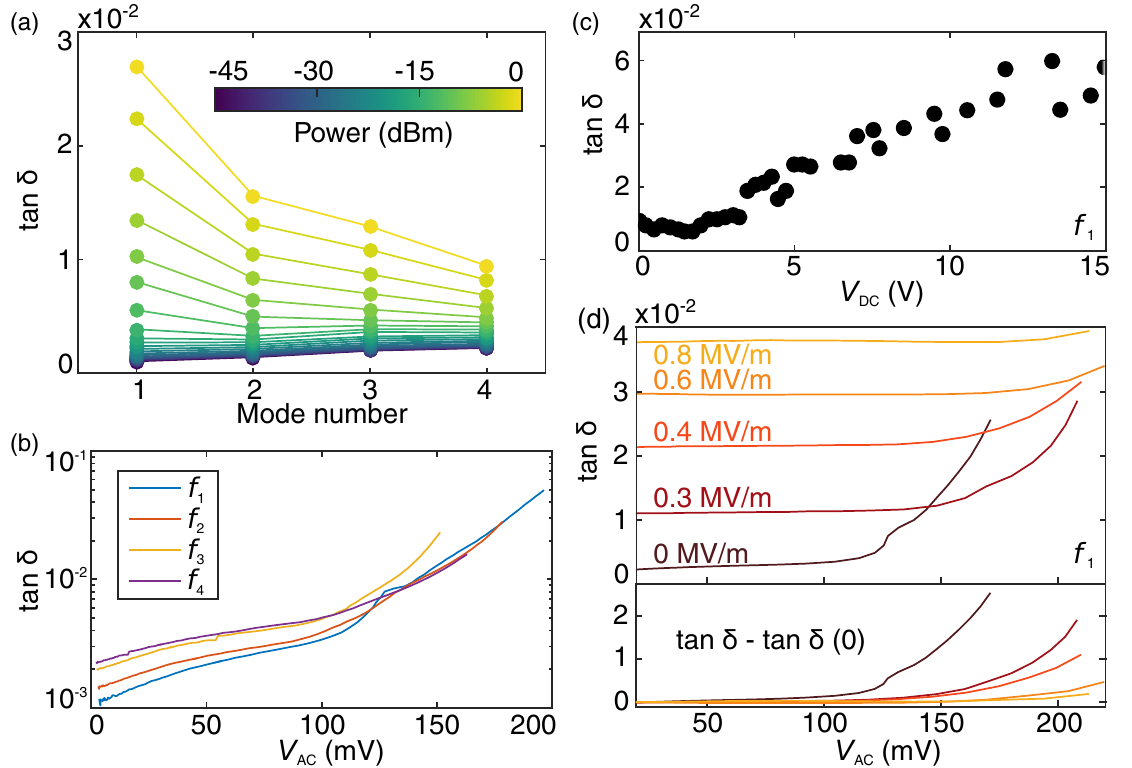}
			\end{minipage}\hfill
			\begin{minipage}[t]{0.22\textwidth}
				\caption{\label{fig:Fig3}
					Dielectric losses in \ce{SrTiO3} probed by the CPW resonator.
					(a) $\tan\,\delta$ vs mode number at zero dc field measured at different rf powers (curves are 2 dBm apart).
					(b) $\tan\delta$ vs the ac voltage inside the cavity at resonance for the first four modes.
					(c) $\tan\,\delta$ vs dc voltage for the first mode of the resonator probed at fixed rf power.
					(d) Loss tangent for the first mode of the resonatoras a function of the internal ac voltage for different values of the dc voltage. The bottom panel shows the overlapped curves, obtained by subtracting the value of $\tan\,\delta$ at minimum $V_\mathrm{AC}$.
				}
			\end{minipage}
		}
	\end{minipage}
	\end{figure*}
	
	\section{POWER AND FIELD DEPENDENCE OF MICROWAVE LOSSES}
		
	\indent \indent Due to their low intrinsic losses~\cite{Singh2014}, superconducting MoRe resonators provide a good platform for characterization of the loss mechanisms of the surrounding dielectric environment. Dielectric losses are typically evaluated through the loss tangent of the substrate ($\mathrm{tan\, \delta}$) defined as the ratio between the imaginary ($\varepsilon''$) and real part ($\varepsilon'$) of the relative permittivity:
	\begin{equation}
	\mathrm{\tan\,\delta \equiv {\varepsilon''}/{\varepsilon'}},
	\end{equation}
	that can be directly obtained from fitting the reflection coefficient from a single port cavity  ($\mathrm{|S_{11}|}$) with its definition:
	\begin{equation}
	\label{eq:S11_ki_ke}
	\mathrm{|S_{11}|} = \frac{2f_0/Q_\mathrm{int} - \Delta f - 4i\pi(f - f_0)}{\Delta f - 4i\pi(f - f_0)},
	\end{equation}
	where $\mathrm{\Delta} f$ is the extracted full-width half-maximum of the resonance peak, $Q_\mathrm{int}$ is the internal quality factor and $f_\mathrm{0}$ its frequency. The loss tangent of the resonator ($1/Q_\mathrm{int}$) usually incorporates both surface losses (originating from amorphous interfaces) and substrate losses. By comparison with the identical resonator fabricated on top of sapphire we can estimate the contribution of the radiation and quasiparticle losses to be less than $\mathrm{7\cdot10^{-4}}$. We therefore use 1/$Q_\mathrm{int}$ and tan $\delta$ interchangeably.

	Fig.~\ref{fig:Fig3}(a) shows the loss tangent for the first four resonance modes for different input rf powers ($P_\mathrm{in}$). Each curve corresponds to a different value of the rf power. For low values of $P_\mathrm{in}$ ($P_\mathrm{in} < 10$ dBm), $\tan\delta$ for all modes is weakly affected by the input power and it increases with mode number. When increasing the input rf power, the losses become strongly affected by the ac field and their mode dependence is reversed, as lower modes seem to react more strongly to the input rf power.
	
	In Fig.~\ref{fig:Fig3}(b) we plot tan $\delta$ of the first four modes, but now as a function of the internal ac voltage rather than input power. This is important because the different internal quality factors of the modes result in different amplitudes of the ac voltage inside the resonator for the same input power. At low ac-voltages ($V_\mathrm{AC} < 90 $ mV), we observe a monotonic dependence of tan $\delta$ on mode number: lower modes exhibit lower tan $\delta$. This suggests that tan $\delta$ is frequency dependent, similar to what was observed in~\cite{Geyer2005} for gigahertz frequencies. In this range, all modes show a linear dependence on the ac voltage with similar slopes (see Supplemental Material S4). For higher amplitudes of the ac voltage ($V_\mathrm{AC} > 90 $ mV) we find an exponential increase of the losses for all four modes. In the high power region, the peaks become progressively broader and we also see signatures of nonlinear damping. Details on the derivations and fits of the data are presented in Supplemental Material S4.

	The observed power dependence of tan~$\delta$ is not consistent with the typical behavior of dissipation in the two-level system (TLS) model, a common source of losses at the metal-substrate/metal-air interfaces, since such interfacial losses usually show an opposite saturation behavior~\cite{Gao2008, Wenner2011}. This suggests that the measured $\tan\delta$ likely originates from substrate losses.
	
	To better understand the loss mechanisms, in Fig.~\ref{fig:Fig3}(c) we also investigate the dependence of $\tan\,\delta$ as a function of dc electric field extracted from the resonance measurements for the first dc sweep. The increase of $\tan\,\delta$ with $V_\mathrm{DC}$ is in agreement with the so-called quasi-Debye mechanism~\cite{Tagantsev2000,Vendik2002}, which is the dominant dissipation channel in incipient ferroelectrics. This mechanism originates from the interplay between dc and ac components of the excitation. The dc field breaks the lattice symmetry, making the phonon spectrum field-dependent. The ac field acts as a time-modulation of the phonon frequencies, which are continuously driven out of equilibrium with consequent energy dissipation.
	This increase has been already observed in \ce{SrTiO3} single crystals~\cite{Findikoglu1999c}, while it is in striking difference with several studies on thin films, where the application of an electric field leads to a decrease in  $\tan\,\delta$~\cite{Findikoglu1995, Findikoglu1999b}.

	Fig.~\ref{fig:Fig3}(d) shows the ac-induced losses for different values of the dc voltage for the first resonance mode where, similarly to Fig.~\ref{fig:Fig3}(a,b), two different regimes can be distinguished. It is possible to compare the losses related to the ac field at different dc bias by considering their rescaled values: $\tan\delta(V_\mathrm{AC})-\tan\delta(V_\mathrm{AC} = 0)$, which are plotted in the bottom panel of Fig.~\ref{fig:Fig3}(d). The presence of a dc field reduces the magnitude of the ac-induced losses, which suggests that the two dissipation channels are not fully independent. 
	
	In contrast to what is observed in thin films, we measured an increase of the losses with the magnitude of both ac and dc fields. According to Zubko and Vasil'ev~\cite{Zubko2009}, the ac and dc field dependence of the losses is determined to a large extent by the parameter $\xi_\mathrm{S}$, which characterizes the quality of the crystal ($\xi_\mathrm{S} \approx 0.02 - 0.06$ for high-quality single-crystal \ce{SrTiO3} and $\xi_\mathrm{S} > 0.5$ for thin films~\cite{Vendik1997,Geyer2005}). For values of $\xi_\mathrm{S} < 0.25$, tan~$\delta$ should monotonically increase with both dc and ac electric fields, which is in agreement with our measurements.

	\begin{figure*}[ht]
	\begin{minipage}[c]{\textwidth}
		{
			\begin{minipage}[t]{0.68\textwidth}
				\mbox{}\\[-\baselineskip]
				\includegraphics[width=\linewidth]{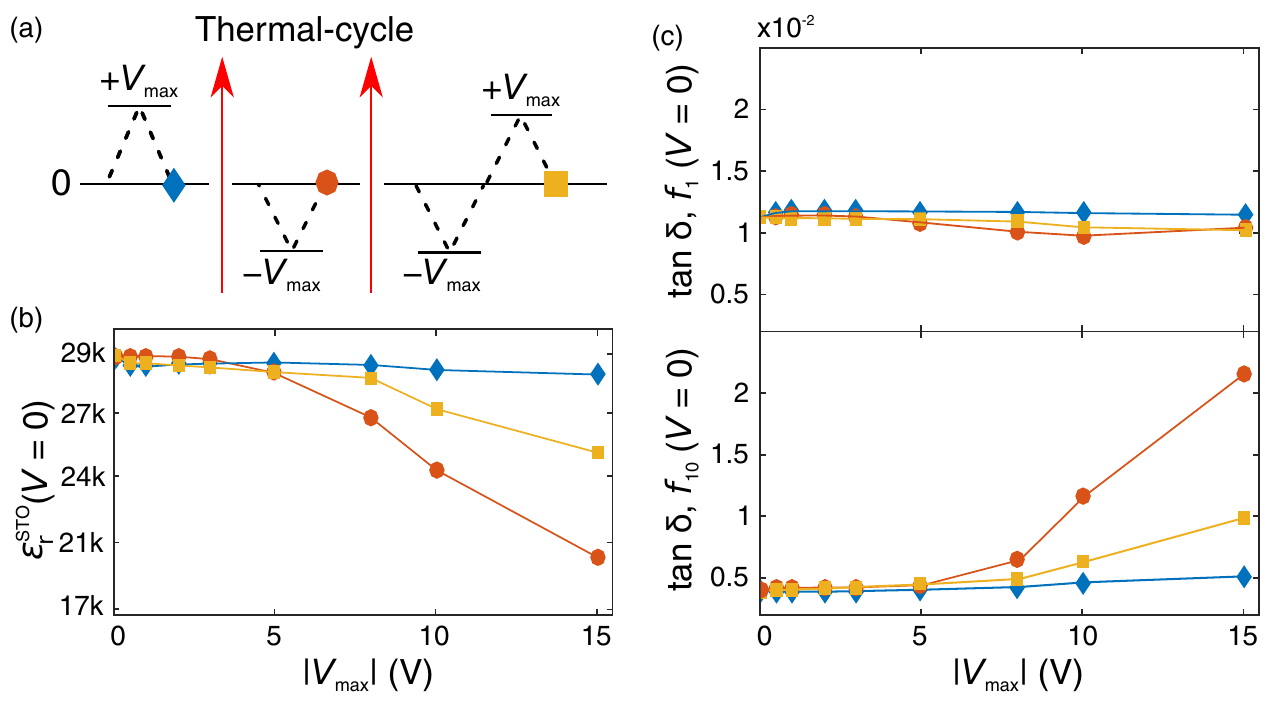}
			\end{minipage}\hfill
			\begin{minipage}[t]{0.29\textwidth}
				\caption{ Hysteretic effects in the dielectric response of \ce{SrTiO3}. 
					(a) Diagram showing the biasing and measurement sequence used to characterize the dependence of $\varepsilon^\mathrm{STO}_r$ (b) and $\tan\,\delta$ (c) on the field polarity. Measurements are acquired at zero dc bias after the application of a given voltage difference ($V_\mathrm{max}$) between the line and the ground plane. We kept a constant polarity for the different $V_\mathrm{max}$ values. After reaching $V_\mathrm{max}=\pm$ 15 V, a thermal cycle to 300 K restores the initial conditions of the sample. (Top: fundamental mode (\ce{f1}), bottom: mode \ce{f10}.)} \label{fig:Fig4}
			\end{minipage}
		}
	\end{minipage}
	\end{figure*}
	
	\section{DEPENDENCE OF $\varepsilon_r$ AND $\tan\,\delta$ ON FIELD POLARITY}
	
	\indent\indent In Fig.~\ref{fig:Fig4}, we explore the influence of voltage polarity on the hysteretic response of the dielectric properties of \ce{SrTiO3}. To do so, we followed the measurement sequences sketched in Fig.~\ref{fig:Fig4}(a): starting from zero bias, we applied a voltage ramp up to a target value $V_\mathrm{max}$ and then back to $V = $ 0, where the CPW resonator spectrum is acquired. This procedure is repeated for increasing values of $V_\mathrm{max}$ up to 15 V, always with the same polarity. We then perform a thermal cycle up to room temperature, while keeping the line grounded, to restore the initial sample conditions. We probed under three biasing schemes: positive, negative and positive after negative (zero average). 
	
	Fig.~\ref{fig:Fig4}(b) shows that the value of $\varepsilon_r$ at zero field is not symmetric with respect to the applied field polarity and, in particular, it is significantly affected only by negative biases. We thus link the origin of the reduction of permittivity to the presence of oxygen vacancies, since a negative bias can increase their concentration in the surroundings of the waveguide, where the dielectric environment is probed. This scenario is supported by the partial recovery observed when a positive voltage follows a negative one (yellow squares of Fig.~\ref{fig:Fig4}). This explains the low tunability reported for CPW resonators realized on top of \ce{SrTiO3} thin films; since even the very low defects concentration present at the surface of our substrate is already capable of lowering the dielectric response by more than 25 \% after the application of 15 V. Even though the cooldown conditions were kept unchanged, the values of $\epsilon_r$ and tan~$\delta$ at zero bias upon cooldown show a spread of about 2 \%. This value is negligible with respect to the observed variations and may be caused by different stable configurations of the oxygen vacancies at room temperature.
	This asymmetric response is also observed in the dielectric losses presented in Fig.~\ref{fig:Fig4}(c). However, we note that the measured response depends unexpectedly on the mode number. As a general trend, a monotonic increase of the losses is observed, which is explained by a progressive accumulation of polarized defects for negative biases~\cite{Vendik2002}. However, for low mode numbers and bias values below 10 V, the application of an electric field results in a decrease of $\tan\,\delta$. In the Supplemental Material S5 and S6, we provide an extensive mapping of the losses showing the evolution of $\tan\,\delta$ as a function of the maximum applied voltage, mode number and input power. In particular, upon having applied a dc voltage of -15 V and performing the same analysis as in Fig. \ref{fig:Fig3} (a), we observe a stronger dependence of $\tan\delta$ on mode number at low powers, whereas, for higher input powers,  $\tan\delta$ of the first four modes shows a similar behavior.
	
	We also extract the polarizability parameter $E_0$ and the horizontal shift of the maximum of $\varepsilon^\mathrm{STO}_r$ by fitting the $\varepsilon^\mathrm{STO}_r$ vs. $V_\mathrm{DC}$ curves during the voltage sweeps (Supplemental Material S7). Finally, to support our interpretation of the fatigue mechanism, in Fig. S8 of the Supplemental Material, we perform a measurement of the time evolution of $\varepsilon_r^\mathrm{STO}$ upon applying a small negative dc voltage. The time response of the dielectric constant indicates the presence of slow relaxation dynamics, which is in accordance with the proposed model of drift of charged defects.
	
	An advantage of the described method is that the superconducting cavity is fabricated directly on top of the \ce{SrTiO3} and serves as a local probe of the underlying dielectric. In addition, the use of a reference sample facilitates the extraction of quantitative information on the dielectric properties of the material. This measurement technique is compatible with a wide variety of complex oxides, it enables local electrostatic gating and it can serve as a platform for time-resolved measurements in the millisecond range.

	\section{CONCLUSIONS}
	
	In conclusion, we realized a superconducting coplanar waveguide resonator directly on top of a single-crystal \ce{SrTiO3}(001) substrate and probed its dielectric properties as a function of dc voltage, rf power injected in the line and mode number. The effective dielectric constant measured in the initial condition was $\sim$ 27,000, lowered below 4,000 by applying a dc voltage of 15 V. The inhomogeneous distribution of the electric filed in the resonator geometry allowed us to uncover an asymmetric response with respect to the field polarity of both $\varepsilon_r$ and $\tan\,\delta$, with a progressive reduction of the resonator's tunability range, pointing towards the presence of charged defects, such as oxygen vacancies, as the origin of the observed behavior. The narrow gap (10 $\mu$m) increases the sensitivity of the microwave response to effects on a lengthscale of the order of the gap size. By further shrinking its dimensions, a microwave cavity can act as a local probe capable of studying, for example, the dynamics and the dielectric behavior of single domain walls, characteristic of \ce{SrTiO3} at low temperatures. Our results demonstrate the sensitivity of superconducting microwave cavities as probes of quantum matter and provide a robust background for future studies on oxide heterostructures probed by superconducting resonators.\\

	\section*{ACKNOWLEDGMENTS}
	The authors would like to thank Daniel Bothner and Mark Jenkins Sanchez for their help during data analysis and Peter G. Steeneken and Teun M. Klapwijk for the fruitful discussions. This work was supported by The Netherlands Organisation for Scientific Research (NWO/OCW) as part of the Frontiers of Nanoscience program and by the Dutch Foundation for Fundamental Research on Matter (FOM). The research leading to these results has received funding from the European Research Council under the European Union's H2020 programme/ERC Grant Agreement n. 677458.

\pagebreak

\onecolumngrid
\appendix
\newpage

\section*{Supplemental Material}
\setcounter{figure}{0}  
\renewcommand{\figurename}{FIG. S\!\!}
\begin{figure}[ht!]
	\includegraphics[width=0.5\linewidth]{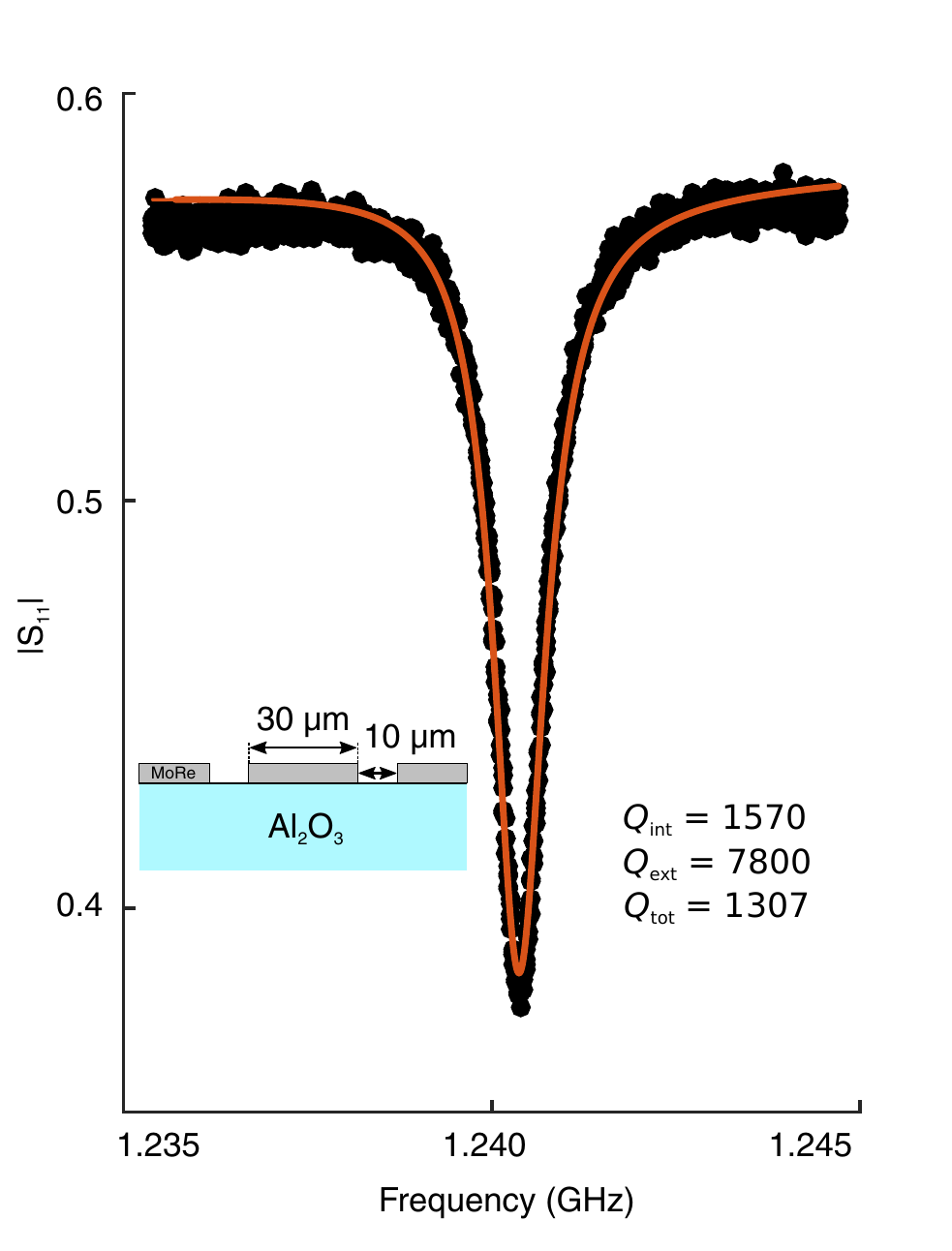}
	\caption{\label{fig:FigS1} First mode of a geometrically identical CPW  resonator fabricated on top of a sapphire substrate.}
\end{figure}

\begin{figure}[ht!]
	\includegraphics[width=0.9\linewidth]{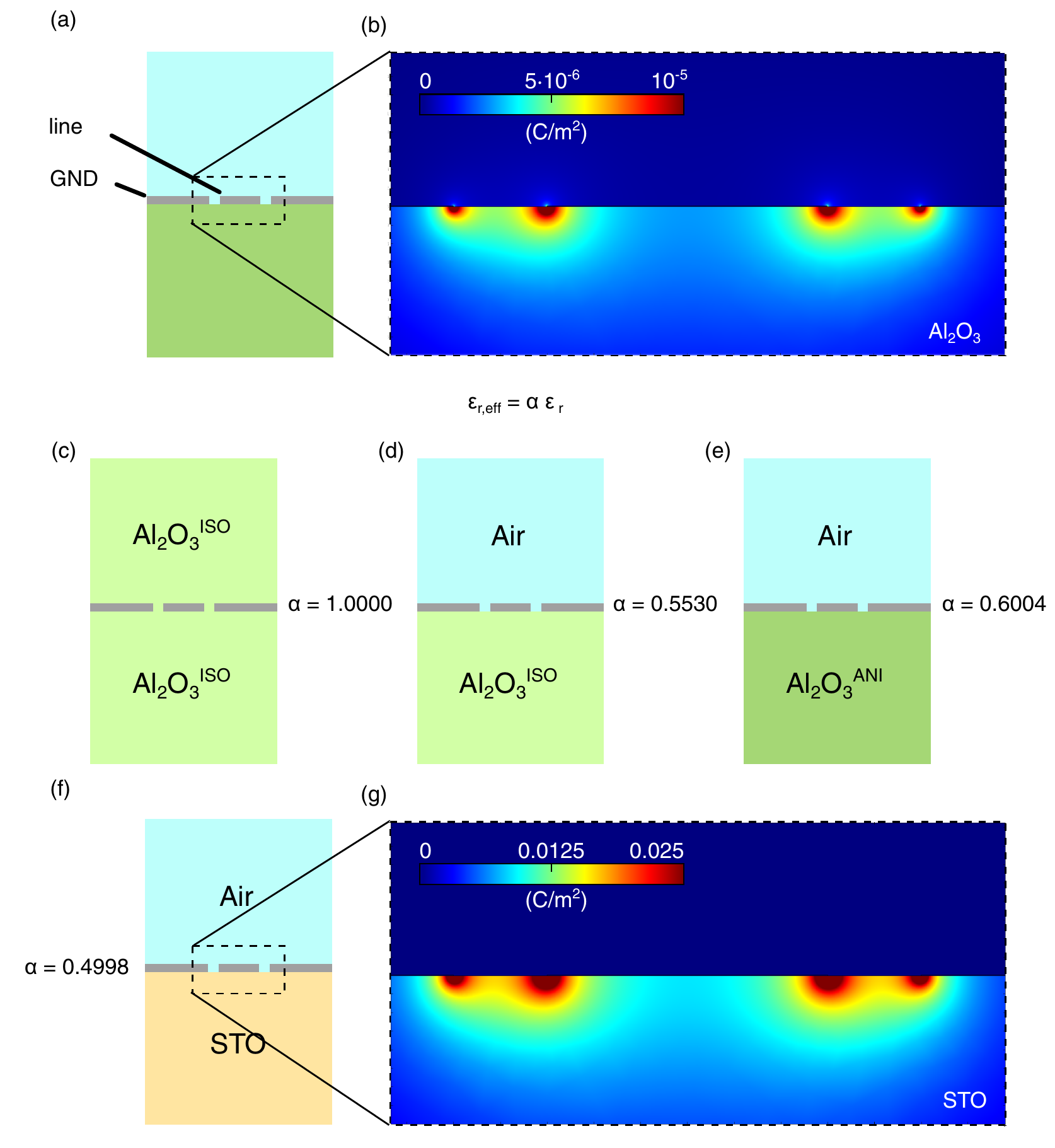}
	\caption{\label{fig:FigS2} Evaluation of the geometrical correction factor $\mathrm{\alpha}$ through finite elements simulations.}
\end{figure}

Fig. S\ref{fig:FigS1} shows a reflection measurement of a geometrically identical CPW resonator fabricated on top of \ce{Al2O3} (sapphire). The fundamental resonance frequency of the sapphire sample is 1.241 GHz and has been used as a reference to extract the relative dielectric constant of the \ce{SrTiO3} (see main text and Fig. S\ref{fig:FigS2}). The extracted loss tangent of the resonator fabricated on sapphire was tan $\delta = 1/Q_{int} = 6.3 \cdot 10^{-4}$, which sets an upper bound for the quasiparticle and radiation losses in a superconducting cavity of the given geometry. The tan~$\delta $ of the (identical) resonator fabricated on top of \ce{SrTiO3}, on the other hand, was measured to be in the order of tan $\delta \approx 10^{-2}$, which can therefore be attributed almost entirely to losses in the \ce{SrTiO3}.

\newpage

We calculated the effective dielectric constant ($\varepsilon_{r,eff} = \alpha \varepsilon_r$) of \ce{Al2O3} and \ce{SrTiO3} using a finite elements model (Fig. S\ref{fig:FigS2}). A cross-sectional schematic of the modeled geometry is presented in (a). The central line is 30 $\mu m $ wide and the spacing is 10 $\mu m$. The resonator is taken to be embedded between two dielectric regions, each of which 500 $\mu m$ thick. By comparing the capacitance of the resonator completely embedded in the dielectric (c) and the one where the top dielectric is air (d,e,f), we can extract the ratio $\alpha$, which is typically around 0.5. In (b) we show a finite elements simulation of the displacement field calculated for case (e), where the anisotropy of the dielectric constant of \ce{Al2O3} is taken into account. For the sapphire sample we considered three different cases: a  resonator completely embedded into an isotropic \ce{Al2O3} volume (c), a resonator fabricated on top of an isotropic \ce{Al2O3} substrate (d), a resonator fabricated on top of an anisotropic \ce{Al2O3} substrate (e). The same simulation is carried out for the case of an isotropic \ce{SrTiO3} substrate (f), the corresponding distribution of the calculated displacement field is presented in the panel (g). The dielectric constant for the isotropic \ce{Al2O3} cases were considered as $\varepsilon_r^{Al_2O_3} = 9.395$. (c) is used as a reference value for calculating $\alpha$ in (d,e).
\newpage

\begin{figure}[ht!]
	\includegraphics[width=0.75\linewidth]{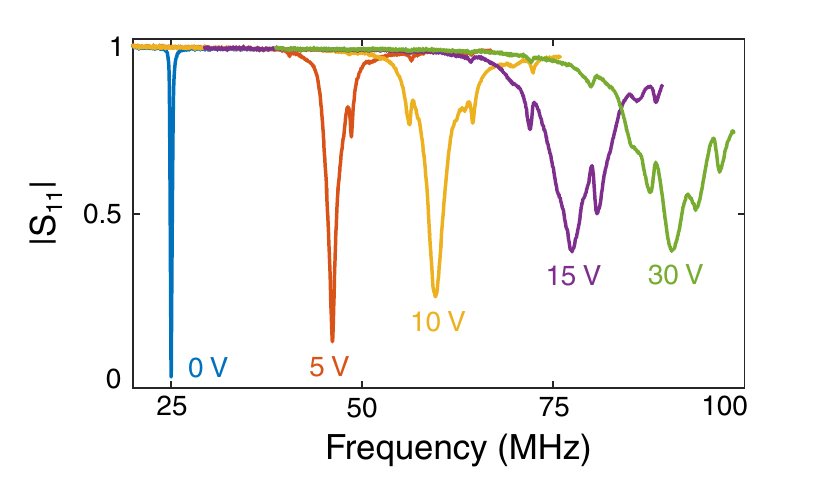}
	\caption{\label{fig:FigS3} First cavity mode for the CPW resonator on top of the \ce{SrTiO3} substrate as a function of the dc voltage.}
\end{figure}	
Fig. S\ref{fig:FigS3} shows the spectrum of the first cavity mode of the CPW resonator on top of the \ce{SrTiO3} sample as a function of the dc voltage. Upon applying a voltage of $V_\mathrm{DC} =$ 30 V, the frequency the mode increases by about 260$\%$. The side peaks appearing at higher dc voltages arise from slot modes on the chip due to non-perfect grounding and becomes visible because of the progressive lowering of the Q-factor of the resonator under dc bias (see main text).

\newpage

\begin{figure}[ht!]
	\includegraphics[width=1\linewidth]{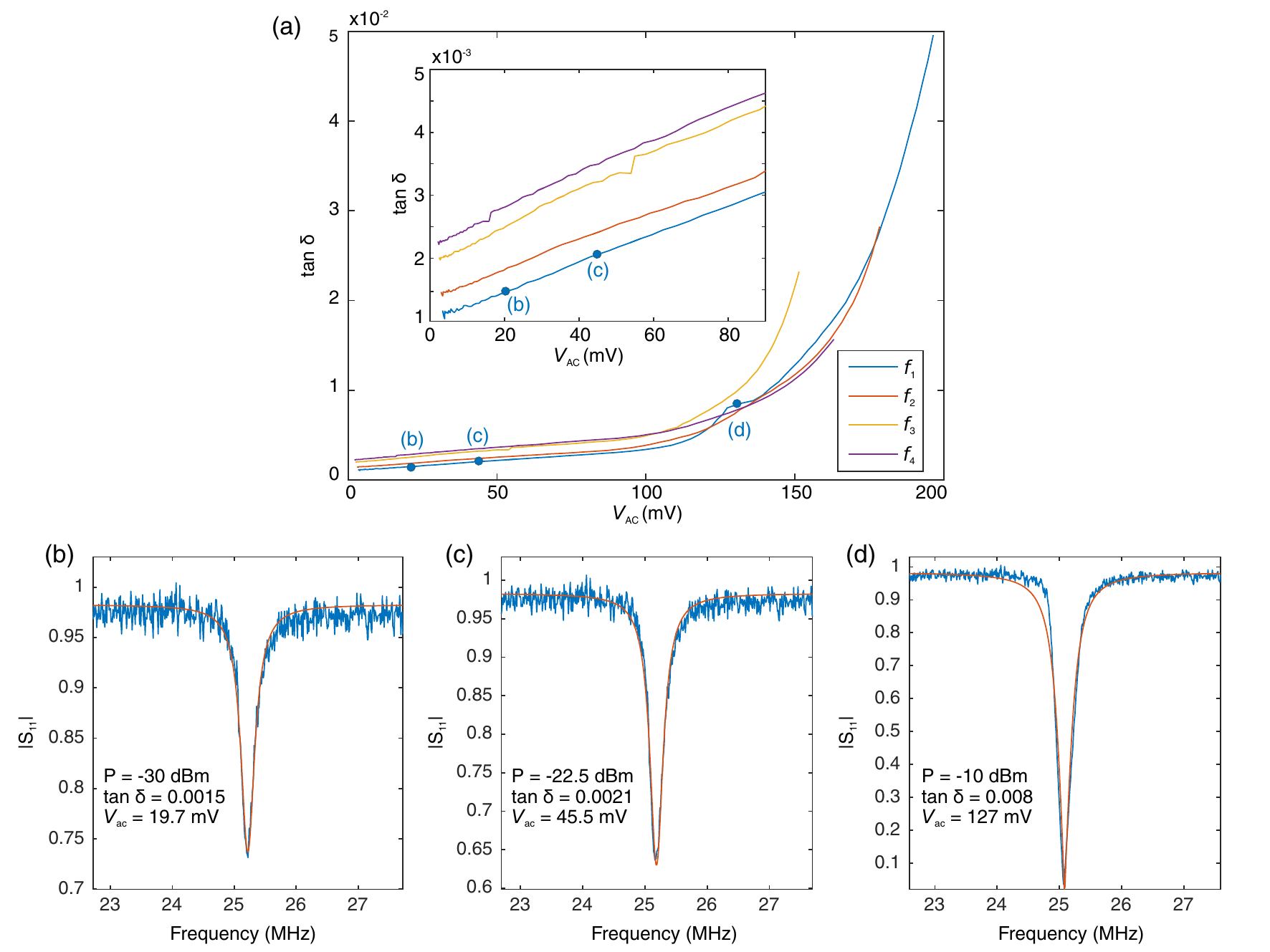}
	\caption{\label{fig:FigS4} (a) tan $\delta$ vs. the ac voltage inside the cavity for the first four modes plotted on a linear scale. Inset: zoom-in of the traces for low voltages  ($V_\mathrm{AC}<$90~mV).  (b-d) Traces and fits of the fundamental mode (blue curve in (a)) at three different input powers (indicated with the blue dots in (a)) taken at zero dc field.}
\end{figure}	

Fig. S\ref{fig:FigS4}(a) shows $\tan\delta$ of the first four modes as a function of the ac voltage inside the cavity (same as Fig. 3(a) of the main text) plotted on a linear scale. Two regimes can be distinguished: for low ac voltages, all curves show a linear dependence with similar slopes (inset of Fig. S4(a)); for voltage amplitudes above 90 mV, the losses start increasing exponentially with ac field.

In Fig. S\ref{fig:FigS4} (b-d) we show the traces and the fits of the fundamental mode at taken at three values for the ac voltage denoted by the blue points (b-d) in Fig. S4(a). The fitting is done using:
\begin{equation}
\label{eq:eq1}
|S_{11}| = \frac{\frac{\kappa_{i} - \kappa_{e}}{2} - i(f - f_1)}{\frac{\kappa_{i} + \kappa_{e}}{2} - i(f - f_1)}
\end{equation}
where $\kappa_i$ and $\kappa_e$ are the internal and external dissipation rates of the cavity (in Hz) and $f_1$ is the resonance frequency of the first mode. $\tan \delta$ is then defined as:
\begin{equation}
\label{eq:eq2}
\tan\delta = \frac{\kappa_i}{f_1}
\end{equation}
We note that eq. 4 from the main text can be obtained by substituting eq. \ref{eq:eq2} into eq. \ref{eq:eq1} (considering $\kappa_i + \kappa_e = \Delta f$). When the internal and external dissipation rates are equal, the resonator is said to be critically coupled and $|S_{11}|$ becomes zero at resonance (Fig. S\ref{fig:FigS4} (d)). For ac voltage amplitudes above 120 mV the traces start to show a sign of nonlinear damping which introduces a systematic error to the fitting.

Using the fitted values, one can obtain the total energy stored in the cavity at resonance as:

\begin{equation}
\label{eq:eq3}
|A_0|^2 = \frac{2}{\pi}P_\mathrm{in}\frac{\kappa_e}{(\kappa_i+\kappa_e)^2+4(f-f_1)^2}
\end{equation}
The energy, in turn, is equal to the electric potential energy in the resonator:		\begin{equation}
\label{eq:eq3}
|A_0|^2 = U_E = \frac{l}{4}C^{'}V_{AC}^2
\end{equation}	
where $V_\mathrm{AC}$ is the voltage amplitude at the electric field antinode inside the resonator, $C'$ is the capacitance per unit length (for our case $C' = 434~\mathrm{nF/m}$  at $V_\mathrm{DC} = 0 \,\mathrm{V}$) and $l$ is the resonator length. The ac voltage inside the resonator is then given by:
\begin{equation}
\label{eq:eq4}
V_\mathrm{AC} = \frac{8P_\mathrm{in}}{ lC' \pi}\frac{\kappa_e}{(\kappa_i+\kappa_e)^2+4(f-f_1)^2}
\end{equation}

\newpage

\begin{figure}[ht!]
	\includegraphics[width=0.95\linewidth]{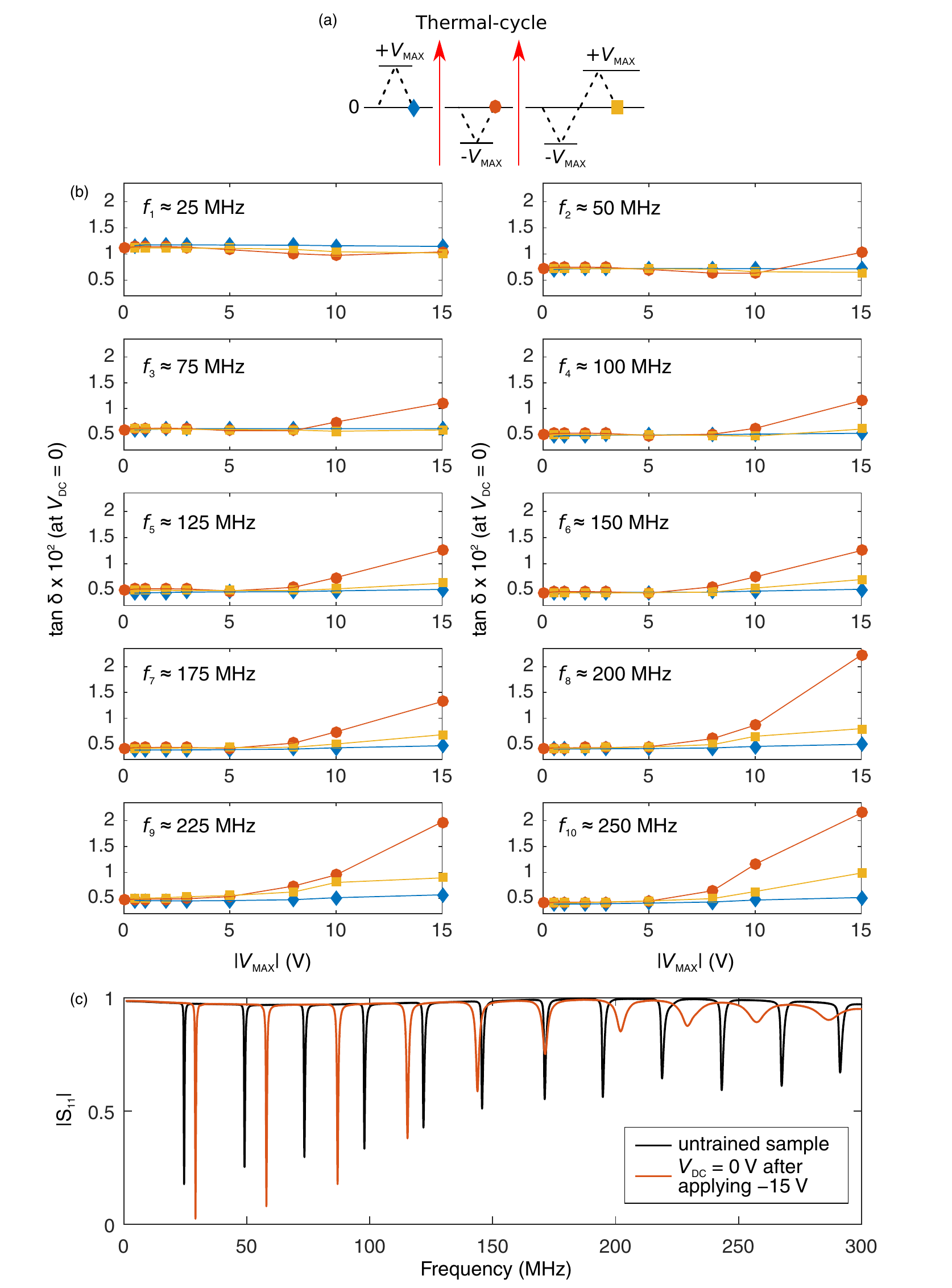}
	\caption{\label{fig:FigS5} Hysteretic effects in the dielectric response of \ce{SrTiO3} for all modes.}		
\end{figure}

In Fig. S\ref{fig:FigS5} we report the hysteretic response of the tan~$\delta$ of \ce{SrTiO3} for the first ten cavity modes of the resonator. The diagram in (a) illustrates the biasing and measurement sequence used to characterize the dependence of tan $\delta$ at $V_\mathrm{DC} = 0$ vs the applied field polarity (see main text). In (b) we present tan~$\delta$ as a function of maximum applied voltage for the first ten resonance modes. In (c) there is the comparison of the reflection spectra taken immediately upon cooldown (black) and after having applied -15 V to the sample (red), qualitatively showing a broadening of the higher modes (increase of losses).
\\\\
\newpage
\begin{figure}[ht!]
	\includegraphics[width=0.75\linewidth]{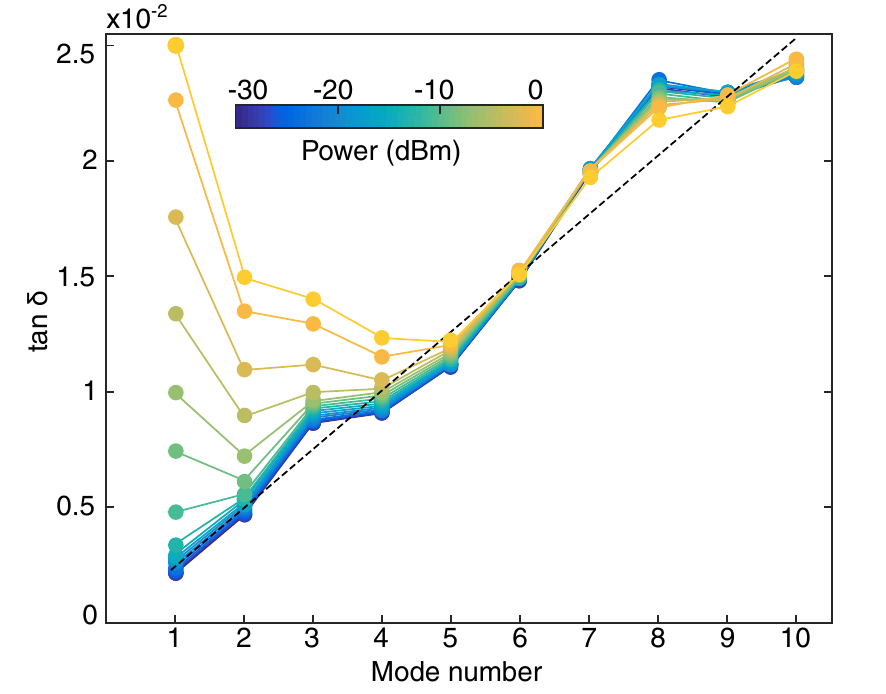}
	\caption{\label{fig:FigS6} $\tan\,\delta$ of a trained sample (after having applied $V_\mathrm{DC}$ = -15 V) vs mode number at zero dc field measured at different rf powers (curves are 2 dBm apart).}
\end{figure}
Fig. S\ref{fig:FigS6} shows the losses of the resonator as a function of input power for the first ten modes after having applied $V_\mathrm{DC}$ = -15 V. At low powers, $\tan\,\delta$ shows a linear dependence on mode number (blacked dashed line represents a linear fit). At powers above $\approx$~10~dBm, the $\tan\,\delta$ of the lower modes (1-4) starts to increase abruptly, similarly to Fig. 3 (a) in the main text. It is important to note that the $\tan\delta$ of the higher modes at low rf power significantly increases due to fatigue of the substrate. Upon performing a thermal cycle, the losses are restored to their initial values.	

\newpage
\begin{figure}[ht!]
	\includegraphics[width=0.95\linewidth]{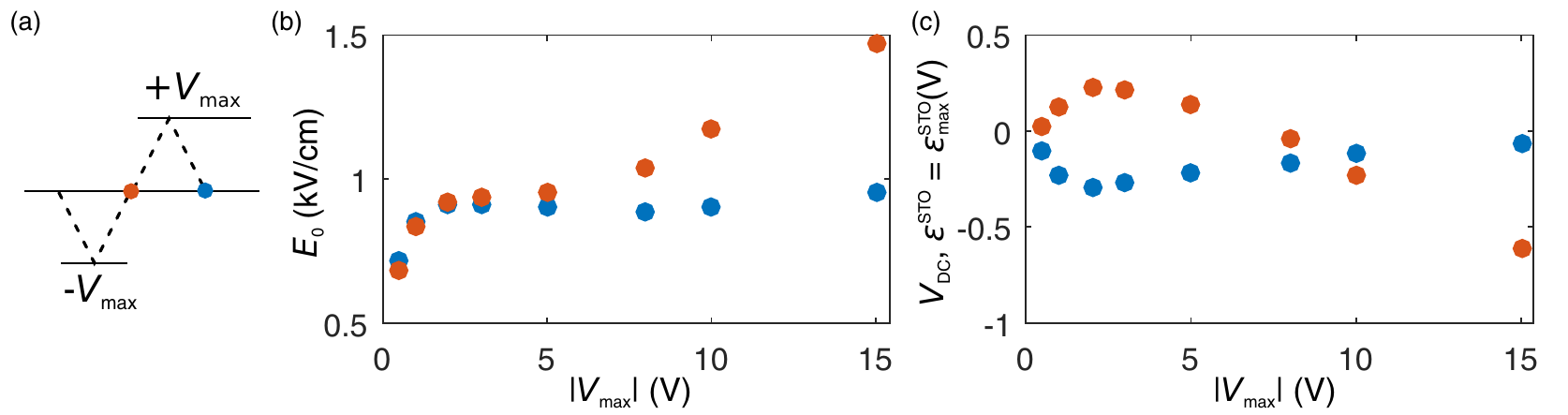}
	\caption{\label{fig:FigS7} Hysteresis of the polarizability parameter and position of the maximum of the dielectric constant as a function of the maximum dc field applied.}
\end{figure}

Fig. S\ref{fig:FigS7} shows the extracted polarizability parameter $E_0$ (see main text) and horizontal shift of the maximum of the dielectric constant extracted from the $\varepsilon^\mathrm{STO}_r$ vs. $V_\mathrm{DC}$ curves. The biasing and measurement sequence is illustrated in the diagram in (a). The red points correspond to the $-V_\mathrm{max}\leqslant V\leqslant 0$ and the blue points correspond to the $V_\mathrm{max}\geqslant V\geqslant 0$ sweeps. The value of $E_0$ as a function of maximum applied voltage is reported in (b). Negative voltages seem to decrease the polarizability of the substrate, the application of a positive voltage determines a partial recovery of the polarizability. (c) shows the horizontal shift of the maximum in the $\varepsilon^\mathrm{STO}_r$ vs $V_\mathrm{DC}$ curves as a function of the maximum applied voltage. We interpret this slight hysteresis as an indication of a small ferroelectric response of the substrate arising from the low-density oxygen vacancies.

\newpage

\begin{figure}[t]
	\includegraphics[width=0.75\linewidth]{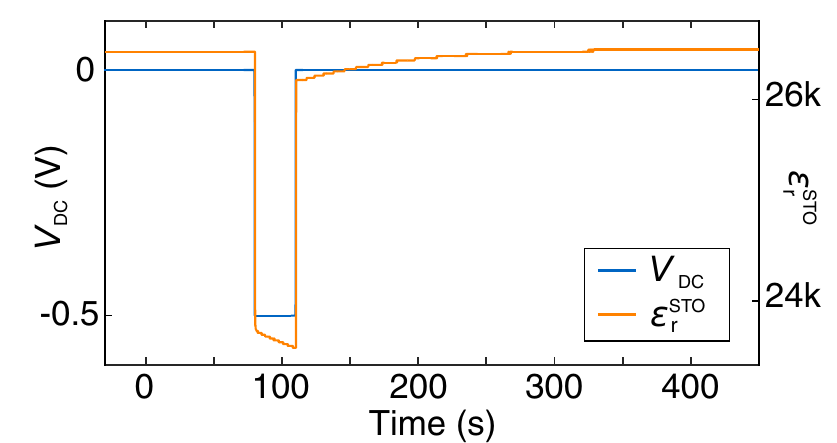}
	\caption{\label{fig:FigS8}
		Time plot of the evolution of $\varepsilon_r$ of the \ce{SrTiO3} (orange line) upon the application of a voltage pulse of $\mathrm{- 0.5 V}$ (blue line). $\varepsilon_r$ is monitored by tracking the frequency of the first mode of the resonator.
	}
\end{figure}

Fig. S8 shows the time evolution of $\varepsilon_r^\mathrm{STO}$ (orange line) upon applying a small dc voltage of $\mathrm{-0.5}$~V (blue line). Here the dc voltage is kept low to avoid significant changes of the resonance frequency, enabling faster data acquisition. The dielectric constant of \ce{SrTiO3} is monitored over time by tracking the frequency of the first cavity mode of the CPW resonator, using a high bandwidth filter of 10 kHz (50 ms for a full spectrum). When a dc field is applied, the dielectric constant drops abruptly, in agreement with Fig.~2(b) from the main text, followed by a slow drift. The dc field is maintained for 30 s and, upon setting the voltage back to zero, $\varepsilon_r^\mathrm{STO}$ exhibits a partial recovery, followed by a slow relaxation to its initial value. In both cases the timescales of the relaxation processes are of the order of seconds. The observed time response is also compatible with processes other than oxygen vacancies, such as relaxation of ferroelastic domain walls. However, the asymmetric response to the field polarity shown in Fig. 4 (main text) points towards the presence of charged defects showing slow drift dynamics, making the presence of oxygen vacancies the most plausible explanation for the observed behavior.

\end{document}